# Robust Contract Evolution in a TypeSafe MicroService Architectures

João Costa Seco[a], Paulo Ferreira[b], Hugo Lourenço[b], Carla Ferreira[a], and Lúcio Ferrão[b]

a NOVA LINCS, Faculdade de Ciências e Tecnologia, Universidade NOVA de Lisboa, Portugal

b *OutSystems*, Portugal

**Abstract** Microservice architectures allow for short deployment cycles and immediate effects, but offer no safety mechanisms for service contracts when they need to be changed. Maintaining the soundness of microservice architectures is an error-prone task that is only accessible to the most disciplined development teams. The strategy to evolve a producer service without disrupting its consumers is often to maintain multiple versions of the same interface and dealing with an explicitly managed handoff period and its inherent disadvantages.

We present a microservice management system that statically verifies service interface signatures against their references and supports the seamless evolution of compatible interfaces. We define a compatibility relation□on types that captures real evolution patterns and embodies known good practices on the evolution of interfaces. Namely, we allow for addition, removal, and renaming of data fields of a producer module without breaking or needing to upgrade consumer services. The evolution of interfaces is supported by runtime generated proxy components that dynamically adapt data exchanged between services to match with the statically checked service code.

The model proposed in this paper is instantiated in a core programming language whose semantics is defined by a labelled transition system and a type system that prevents breaking changes from being deployed. Standard soundness results for the core language entail the existence of adapters, and hence ensure the absence of adaptation errors and the correctness of the management model. This adaptive approach allows for gradual deployment of modules, without halting the whole system and avoiding losing or misinterpreting data exchanged between system nodes. Experimental data shows that an average of 57% of deployments that would require adaptation and recompilation are safe under our approach.

## 1 Introduction

Development speed of microservice-based architectures is very attractive from the perspective of agile software development and evolution methodologies. The increased module isolation provided by microservices allows for completely independent tracks of deployment, heterogeneous technological choices, greater team focus, and increased productivity. Microservices mitigate many challenges of the development process of monolithic systems where releasing a new feature or a patch may need to wait for a stable synchronization point in time. Large monolithic applications also have significant system-wide downtime intervals which disappear when using microservices.

In loosely-coupled structures, such as microservice architectures, the effects of deploying new implementations (i.e. versions) are immediate, ideal to support fast changing systems with high availability requirements [2, 5]. Hence, maintenance tasks that only concern one module (e.g. bug fixing, local optimizations, refactoring) can be addressed efficiently, independently, in parallel, and in shorter evolution cycles, thus increasing the overall development and deployment speed. Note, that although microservices can be seen as SOA, they are at the end of the spectrum of SOA with a huge push for loosely-coupled services. As a consequence, microservices avoid relying on a predefined enterprise-service-bus interaction protocol, as is common with *standard* SOA services.

One important downside, that often shadows the benefits of using microservices-based architectures, is the introduction of multiple unchecked failure points. Contract management and service integration become exponentially more difficult with the increasing number of independent services and their interactions [19]. In all modular structures, collaborating parties must agree on a contract or interface to operate on. In service contracts, syntactic or semantic changes are usually silent as services evolve separately and contracts are not usually formally specified or even documented. Without explicit verification of contracts at deployment time the only symptoms that the system is broken are runtime errors or unexpected behaviours. New contract management methods, that include fine-grained ownership models for services, defining responsibilities of teams over modules, are needed to keep everything in sync and evolve seamlessly. Having a common specification for service contracts is paramount, allowing for the safety of service-based architectures to be checked *a priori* [9, 38].

Checking a system as a whole gives little information about the semantics of deployment actions, that define partial changes in system semantics. Pragmatic and successful, yet costly, strategies to evolve microservice architectures rely on keeping multiple versions of the APIs and explicitly adapting and redirecting new service implementations [22, 23]. In this context, sophisticated mechanisms are needed to coordinate service consumers (caller) and service producers (callee) to match versions on both ends, usually distinguished by different URIs (cf. Netflix's Eureka system [28]). Safety of the evolution is a software engineering challenge which would greatly benefit from automatic tools that can deal with trivial situations where changes are harmless but critical under strict policies, and alert the developers when unsafe deployments are requested. We present a microservice management system that is based on statically checked service interfaces, a flexible exchange format (cf. JSON), and the automatic generation of proxy (adapter) components. This system is inspired by integrated lifecycle management systems in high-level development platforms. Our reference examples include implementations of the Open Service Broker API [17], like Kubernetes or Cloud Foundry, the Eureka system [28], and the low-code OutSystems Platform component Service Center [30], that mechanically check and manage the deployment of system components. Such platforms operate mainly on service level meta-information, like module version identifiers and API descriptions, to establish the safety of deployments present in monolithic systems. We build on the observation that a new model for such environments that takes advantage of the generic meta-information on services already in place and uses extended type-based interface information is capable of ensuring the safety of deployment operations.

We develop a language-based approach using a core programming language, a labelled transition system, and a type system to express the static and dynamic semantics of the system. Our type system ensures the safety of the systems' operations given a set of lightweight checks performed at deployment time. The verification of evolution steps comes in line with other works that type and check unforeseen programming steps [31] and reconfiguration actions [34] in running systems. The transitions captured in our model include internal service computations, communication between services, and deployment operations. Our type system is parametric on a signature-based compatibility relation on service contracts, and we foresee that it can be extended to richer semantic contracts.

In our setting, we require that meta-information about the service interfaces is provided by each development environment, and that there is a runtime support system, on every consumer module, for a companion microservice or component that is equipped with adapter code (proxies) for each one of its producer services, intersecting and adapting all the service calls. The proxy creation mechanism uses the available meta-information (deployment label and runtime type information) to automatically update the proxy components when needed, following a specific one-to-one protocol between producer and consumer services. We formally guarantee that modifications to type definitions in a producer module do not have immediate impact on the whole system by requiring a global downtime, while ensuring that no new data is ever lost. For instance, when a consumer and producer services agree on a type definition for an endpoint, provided by a given third module, their existing service implementations continue to work correctly if that type definition evolves, using old versions of the type, without loosing any new data. Default values are used when needed. Notice, that we focus on the evolution of signatures for microservices as these are often tagged as breaking changes and are harder to address. The addition of new services, or the updating of existing ones are base operations in our typed model. Overall, the soundness of the typing discipline formally implies that deployments of "compatible" modules do not disrupt the system.

Our research goal is to determine if it is possible to evolve a microservice architecture while ensuring their soundness and avoiding heavy processes of adaptation and redeploying of services. We show that the automatic adaptation provided by our model makes it possible to avoid costly synchronization interactions between the deployment of different modules. We prioritised the automation of deployments over the use of richer contracts (e.g. , pre- and post-conditions). Such requirements are left for future developments as they are fairly orthogonal to our base approach and would require a non-trivial extension to the compatibility relation and to integrate an SMT solver in our signature checking algorithm.

Our framework is inspired by the programming model of the OutSystems Platform, a low-code platform based on visual languages, for the development of enterprise-grade mobile and web applications that abstracts and manages most of the technical details of deploying and maintaining cloud-based data-centric applications. The results of this paper lay the ground for future versions of the platform that automatically and seamlessly will distribute programming modules using microservices instead of heavyweight web containers.

In summary, the contributions of this paper are the following:

- A model for an integrated lifecycle management system targeting a microservices infrastructure;
- A compatibility relation on service contracts, based on real traces of evolution, that ensures that it is always possible to generate adapter code for values of compatible types;
- An adaptation protocol flexible enough to transfer data between different (but compatible) versions of services without losing data (lemma 6);
- A semantics for a microservice management system whose operations include remote calls between services, deployment and undeployment of services;
- A type based structuring discipline for microservices, and a corresponding lightweight preflight check procedure for deployment operations. The type soundness result (theorem 1) certifies the safety of the deployment operations.

The remainder of the paper is organized as follows. In section 2 we describe the architecture of the system, while section 3 introduces a running example to illustrate our compatibility relation. We next describe how the service-based architecture may evolve in section 4 and how services adapt. In section 5 we describe the core language, followed by its type system in section 6, and deploy semantics in section 7. In section 8 we present our formal results, including the soundness of the type system. We conclude the paper by evaluating the impact of the approach, discussing related work and presenting some final remarks.

## 2 System Architecture and Overview

We design the architecture of our microservice management system as depicted in figure 1. The central component in the system, the *global deployment manager*, establishes the connection between different development environments, each one capable of deploying new versions of several services, and each service potentially using different languages and development technologies. We build our design on some characteristics of the runtime support system:

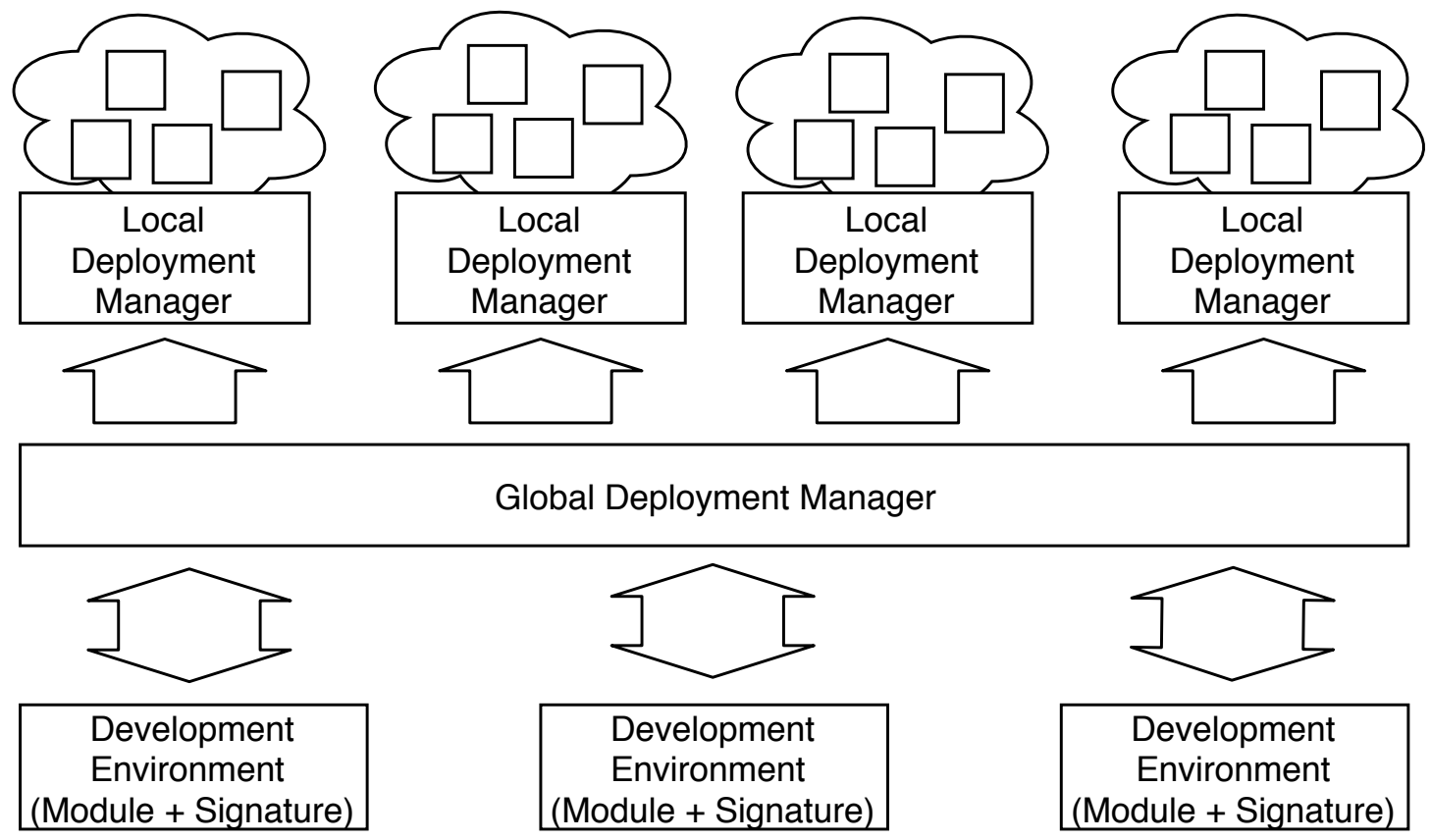

**Figure 1** A centralized control and a distributed runtime

- We assume that microservices are specified using a common contract specification language to express module/service interfaces. This description is a crucial input for the *global deployment manager* and is embodied in our core model by the type description of a service.
- A second assumption to our formal results is that modules fulfil their contracts. This is only possible if developers certify their code against the prescribed contracts. Code certification is achievable at different levels with tests or more formally, using some Hoare-like, logic-based, certification tools.
- A fundamental technical assumption is that we can automatically generate and inject adapter components (proxies) into generic service implementations to transform received and transmitted data. This goal is attainable by manipulating the virtualised images of microservices at the network configuration level and interspersing proxies on every request and response. Proxies are instantiated in our model by the function defined in section 7, figure 8.
- Lastly, to ensure our last result on the preservation of unknown data across services (lemma 6), we require that modules use support for datatypes that preserve unknown fields in the interface, like opaque and unmarshaled data. This data must be preserved in all internal module computations. This is can be achieved by using special data handling layers and libraries, or by using dynamically typed languages like Javascript or Python that treat extra fields in objects blindly. Low-code platforms that compile for mainstream languages and frameworks, such as the OutSystems platform, can easily incorporate this layer into the generated applications. If such mechanism is not available there are no guaranties that data is preserved independently of the data schema and versions used.

Our technical developments include a description language (types) for service interfaces and a single core language to define all module implementations. We then define a compositional property for the system's well-formedness, which corresponds to typechecking all modules against the globally known type signatures of its services. Our semantics is instrumented to deal with unknown values and to generate proxies on demand. In this way we satisfy all the properties above in a core model.

We design the *global deployment manager* to receive requests for the deployment of sets of modules (compilation units) to microservice containers and to check their compatibility with known deployed services as a precondition to their deployment. If a deployment request is deemed compatible with all running services, the operation proceeds in coordination with the local deployment managers, instantiated by extended instances of kubernetes or similar services. Atomicity of deployments must be coordinated by global and local deployment managers. The practical details of this coordination is out of the scope of this paper.

The *global deployment manager* also acts as a registry service (as Eureka system [28]), maintaining meta-information about the signatures of all deployed services (types). If all deployment operations are channelled by such a component, where compatibility is ensured step by step, our system guarantees that no mismatch exists between expected and supplied service interfaces. Deployed microservices communicate amongst each other through said proxies, which are managed by the *global deployment manager*. Services are initially deployed without any proxy installed, as proxies are generated on demand when the first communication between corresponding services happens. We believe that there is a negligible overhead in this lazy instantiation method. Basically, on every communication there is information about the agreed version, and the handshake protocol leads in some cases to a proxy update, whose cost is always attributed to the deployment of one module.

The *local deployment managers* are off-the-shelf components that are the actual executors of deployment actions, coordinating with the *global deployment manager* to achieve the atomicity of deployments.

The *development environments* represent instances of standard tools, that maintain and communicate the known service interfaces in every deployment operation. In our current solution, it's the responsibility of the developer to plan the deployment of groups of modules that are deemed safe. The algorithm that automatically determines the minimal set of services to be updated is a basic analysis on the dependency graph of a modified module. It is out of the focus of this paper and is left for future work.

Our model is instantiated in a core programming language that captures the definition and evolution of a system of communicating microservices. The interface and behaviour of each service is defined by a module. Modules contain a set of definitions (exports) and static references to other modules (imports). Services include a module definition and dynamically generated proxies that bridge the communication between services. The present structure directly maps the underlying structure of OutSystems applications. Nevertheless, it is general enough to be adapted to other technology stacks. In particular, notice that actual services can be implemented with any suitable technology and programming language and that data is exchanged via standard, human-readable, communication formats.

## 3 Running Example

We now present an example of a small application and its evolution that illustrates our model. Consider a catalog of products (figure 2) implemented by three independently

Catalog $V_1$

| | |
|---|---|
| **Definitions** | $Product_{k1} : \{Id_{k2} : \texttt{int}, Name_{k3} : \texttt{string}, Amount_{k4} : \texttt{int}, Discount_{k5} : \texttt{int}\}$ |
| | $Get_{k6} : (id : int) \rightarrow Product = \{Id_{k2} = 1, Name_{k3} = \text{“}HDD\text{”}, Amount_{k4} = 99, Discount_{k5} = 0\}$ |
| | $Save_{k7} : (p : Product) \rightarrow \texttt{string} = \text{“}OK\text{”}$ |

Marketing $V_1$

| | |
|---|---|
| **Definitions** | $Promote_{k8} : (p : Product) \rightarrow Product = p\ \{Name_{k3} = p.Name + \text{“}(Sale)\text{”}, Discount_{k5} = 5\}$ |
| **Reference to Catalog** | $Product_{k1} : \{Id_{k2} : \texttt{int}, Name_{k3} : \texttt{string}, Amount_{k4} : \texttt{int}, Discount_{k5} : \texttt{int}\}$ |

Backoffice $V_1$

| | |
|---|---|
| **Definitions** | $Improve_{k9} : (id : \texttt{int}) \rightarrow \texttt{string} = Save(Promote(Get(id)))$ |
| **Reference to Catalog** | $Product_{k1} : \{Id_{k2} : \texttt{int}, Name_{k3} : \texttt{string}, Amount_{k4} : \texttt{int}, Discount_{k5} : \texttt{int}\}$ |
| | $Get_{k6} : \texttt{int} \rightarrow Product$ |
| | $Save_{k7} : Product \rightarrow \texttt{string}$ |
| **Reference to Marketing** | $Promote_{k8} : Product \rightarrow Product$ |

Catalog $V_2$

| | |
|---|---|
| **Definitions** | $Product_{k1} : \{Id_{k2} : \texttt{int}, Name_{k3} : \texttt{string}, Price_{k4} : \texttt{int}, Discount_{k5} : \texttt{int}, Desc_{k10} : \texttt{string}\}$ |
| | $Get_{k6} : (id : \texttt{int}) \rightarrow Product = \{Id_{k2} = 1, Name_{k3} = \text{“}HDD\text{”}, Price_{k4} = 99, Discount_{k5} = 0, Desc_{k10} = \text{“}2TB\text{”}\}$ |
| | $Save_{k7} : (p : Product) \rightarrow \texttt{string} = \text{“}OK\text{”}$ |

Marketing $V_2$

| | |
|---|---|
| **Definitions** | $Enhance_{k8} : (p : Product) \rightarrow Product = p\ \{Name_{k3} = p.Name + \text{“}(Sale)\text{”}, Discount_{k5} = 5\}$ |
| **Reference to Catalog** | $Product_{k1} : \{Id_{k2} : \texttt{int}, Name_{k3} : \texttt{string}, Amount_{k4} : \texttt{int}, Discount_{k5} : \texttt{int}\}$ |

Catalog $V_3$

| | |
|---|---|
| **Definitions** | $Product_{k1} : \{Id_{k2} : \texttt{int}, Name_{k3} : \texttt{string}, Price_{k4} : \texttt{int},$ ~~$Discount_{k5} : \texttt{int}$~~ $, Desc_{k10} : \texttt{string}\}$ |
| | $Get_{k6} : (id : \texttt{int}) \rightarrow Product = \{Id_{k2} = 1, Name_{k3} = \text{“}HDD\text{”}, Price_{k4} = 99, Desc_{k10} = \text{“}2TB\text{”}\}$ |
| | $Save_{k7} : (p : Product) \rightarrow \texttt{string} = \text{“}OK\text{”}$ |

Marketing $V_3$

| | |
|---|---|
| **Definitions** | $Enhance_{k8} : (p : Product) \rightarrow Product = p\ \{Name_{k3} = p.Name + \text{“}(Sale)\text{”},$ ~~$Discount_{k5} = 5$~~ $\}$ |
| **Reference to Catalog** | $Product_{k1} : \{Id_{k2} : \texttt{int}, Name_{k3} : \texttt{string}, Amount_{k4} : \texttt{int}, Discount_{k5} : \texttt{int}\}$ |

**Figure 2** Evolution of modules Catalog, Marketing, Backoffice.

developed and deployed modules: Catalog, Marketing, and Backoffice. Module Catalog represents a database of products and functions to get and and update products. Module Marketing provides a function to modify a product in order to make it more appealing. Module Backoffice orchestrates the functions of Catalog and Marketing.

Each module is defined by a unique name, a set of type and function definitions (produced by the module), and a set of type and function references (consumed by the module). Definitions exposed by a (producer) module can be used by other (consumer) modules if there is an explicit reference to them in the consumer. Direct references between modules are explicitly maintained by the *global deployment manager* component, in figure 1, and communication between services is performed by the specialized layer that includes the use of dynamically generated proxies.

In our example, module Catalog produces type Product, the datatype of the products stored in a database, and functions Get and Save to fetch and modify products. For simplicity, the functions return fixed values. Exported functions in modules are mapped to service endpoints. Module Marketing contains a reference to type Product from Catalog, and defines function Promote, whose purpose is to automatically improve a product by changing its attributes. In our example the improvement amounts to appending string “ (Sale)” to the product's name and setting a discount. Finally, module Backoffice consumes (has references to) all of exposed elements of other modules, and defines function Improve, that calls functions from other modules. These simple modules establish a triangle of dependencies rich enough to illustrate our deployment discipline. Notice that module Catalog is a producer and does not depend on any other module. Module Marketing is a producer as it exposes the Promote function, and it is also a consumer as it depends on the Product type defined in module Catalog.

Each programming element is labeled with an immutable and unique key, for instance, the key of type Product is $k1$. Editors for low-code visual languages, such as OutSystems or Kony Mobility, manage element keys transparently and automatically, thus there's effectively no overhead for managing element keys.

When the modules are deployed together, we have a *system* consisting of a set of services (in grey in figure 2), where each *service* is running an instance of each one of the deployed modules. They are tagged with version 1 and all references between them are consistent with their related definitions.

In a second phase, we proceed to independently develop version 2 of modules Catalog and Marketing (figure 2). In module Catalog, we modify type Product by renaming Amount to Price, and adding attribute Desc, and in module Marketing, function Promote is renamed to Enhance. Module BackOffice is not modified at this stage, and will become out of sync with the remaining modules. The new and modified elements are highlighted in red and blue, respectively, in figure 2.

In a third phase, version 3 of Catalog is created where attribute Discount of Product has been removed (figure 2). This version cannot be deployed in a system with version 2 of Marketing as function Enhance uses the attribute. If we modify the definition of Enhance by removing the usage of attribute Discount, resulting in version 3 of Marketing, then we can deploy both modules together. We could also deploy version 3 of Marketing first and then version 3 of Catalog, but not the other way around. Moreover, we're allowed to keep attribute Discount in the definition of type Product known by Marketing; correctness is ensured as long the attribute is not used. The removed elements are highlighted in yellow in figure 2.

As the several versions of the modules illustrate, our system is able to overcome the differences between interface definitions, using a communication protocol that automatically adapts to the following kind of changes:

- Adding new attributes to a record type. In Catalog the Desc attribute was added, thus data returned by this module will include values for this attribute. Other modules will preserve this data (as unknown attributes) so that no data is lost when crossing services boundaries.
- Renaming functions. In module Marketing the Promote function was renamed to Enhance, which will impact the endpoints exposed by the service. At runtime the proxy is dynamically generated to use the actual endpoint name when issuing calls, thus not requiring service Backoffice to be updated and redeployed.
- Removing unused attributes or functions. Elements that are not being used require no explicit adaptation.

JSON based exchange is not robust enough to handle these kind of changes, rendering them *breaking changes* in a traditional approach to implementing services. For example, renaming functions results in remote URIs change, and modifying the name of an attribute potentially results in data loss. In contrast, in our approach these kind of changes are allowed without breaking the compatibility between modules. For instance, we can safely deploy version 2 of Catalog and Marketing (figure 2), both together or separately, and keep version 1 of Backoffice: even though the name for function Enhance in module Marketing no longer matches the name known by module Backoffice, at runtime we automatically adapt to the new name.

We define the scope of automatic adaptations by a universal compatibility relation on types, presented in definition 1. Changes that cannot be adapted automatically, and would result in inconsistent system states, are detected and prevented before any deployment action is taken. That is the case when removing functions or attributes from a module that are still being referred by other modules. The decision on how to adapt to this kind of change cannot be made automatically. The verification procedure introduced in this paper, represented by a type system, is able to detect such situations.

## 4 System Behaviour

We now focus on the semantics of service interface synchronization. We illustrate this with a call to function Improve of service Backoffice. As a result we will observe how proxies are dynamically created to adapt services through the differences introduced by the subsequent module evolution.

As a design principle, the visible part of all communications (remote calls) are always performed with relation to the producer's definitions. This amounts to always using the REST identifiers and JSON schema defined in the producer module. Thus, it is the consumer module's responsibility to adapt itself to make all (compatible) changes seamless. To achieve this adaptability at runtime, (consumer) services include a module definition, a deployment label, and a proxy for each one of the module's producers. Deployment labels are unique and fresh whenever a new version of a service is deployed. Proxies store a copy of the signatures of the producer module definitions. Signatures define the data transformations necessary to perform a remote call, one that adapts the arguments, establishes the connection to the right endpoint, and transforms the results of the call back to the original format.

Services are initialised with empty proxies and labelled with the consumer's deployment label to all of their producers. Proxies are (re)created on demand whenever a remote call happens between two services and their deployment labels do not match. Proxies are then updated with the current producer service's meta-information, including its deployment label. In our running example, the system obtained by deploying version 1 of Backoffice and version 2 of Catalog and Marketing is depicted in figure 3a. Notice that service Backoffice has empty proxies, both with deployment label $\ell_3$, to services Catalog and Marketing. Notice that module Marketing consumes a type definition from module Catalog but this is not associated to any proxy.

The handshake protocol for a remote call is as follows. When a remote call is executed, the proxy's current deployment label is compared with that of the producer service. This certifies the synchronization state of the services involved. If deployment labels fail to match, the call is rejected and the producer service replies with its current

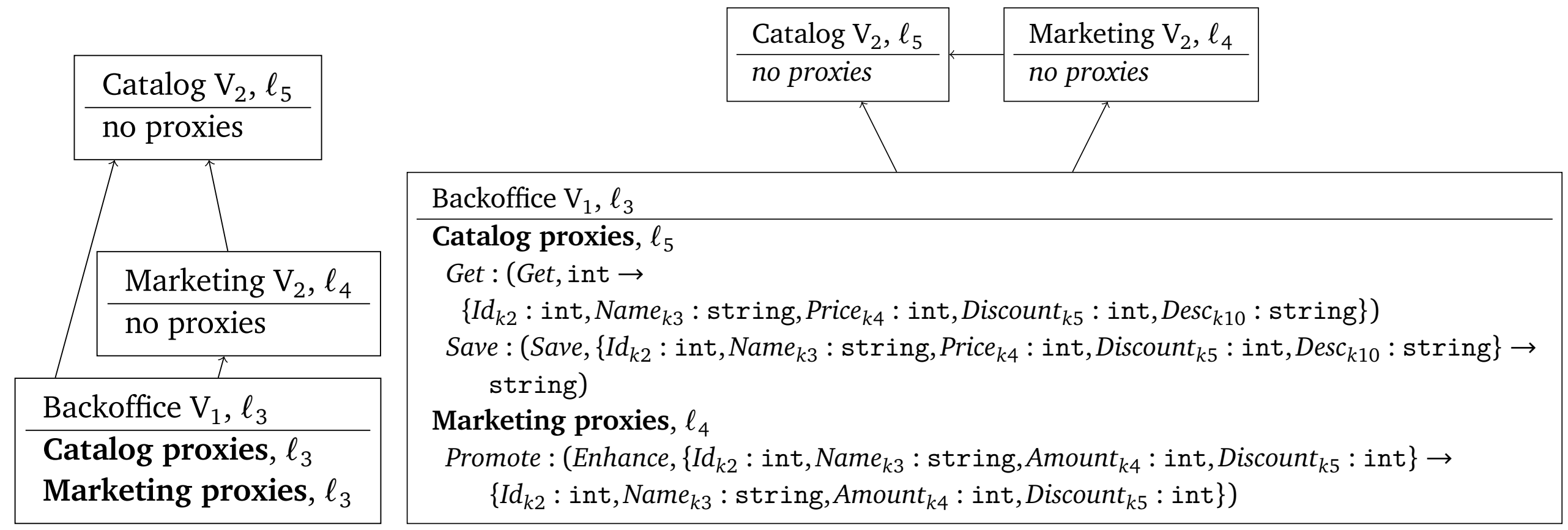

**(a)** System before call. **(b)** System after call.

**Figure 3** Proxies generated as a result of the call Backoffice.Improve.

signature. The consumer service then updates the proxy and retries the remote call. The timeline is shown in figure 4 and the resulting system in figure 3b.

In version 2 of Catalog the definition of record type Product has changed, but all other modules still know and use the old definition. Proxies adapt data to the format expected by the callee: they handle attributes that have been renamed and attributes that are unknown but need to be preserved. Observe in figure 3a that attribute Desc is represented by its key, $k10$, in the communication between services Backoffice and Marketing, whose code was not compiled knowing it. Unknown fields are just passed around the system, and recognised and captured when static information about them is available. Thus, in our approach, services continue to work with older versions of the agreed protocol, preserving data from newer versions if necessary. Any change in one part of the system is not immediately propagated to all involved services, but we ensure that all system communications are safe.

The use of keys goes against the best practices for exchanged messages in REST based architectures, which should always be human-readable [16]. However, in the context of managed microservices of an application where different parties may evolve separately, we need to preserve data for unknown fields which are necessarily tagged with keys instead of names. In our approach the use of keys in messages is transient, it is only necessary while versions are out of sync and data for unknown fields needs to be preserved. Notice that new default values are created in the adapters that add a new field to a datatype. Such values are predetermined for each basic type.

## 5 A core language for services and modules

Our language-based approach is instantiated in a core language, where we represent complete systems, and where we may express and verify the soundness of a system's execution and management operations. We define a static and a dynamic semantics following a syntax-based approach [39]. Our language terminals contain names for

$$
\begin{array}{lll}
e ::= num \mid str \mid e \oplus e & & \textit{(basic values)} \\
\quad \mid f \mid x \mid \lambda x : \tau . e \mid e(e) & & \textit{(lambda calculus)} \\
\quad \mid \{\overline{r_k = e}\} \mid e.r \mid e\,\{\overline{r_k = e}\} & & \textit{(record operations)} \\
\quad \mid t? & & \textit{(waiting thread)}
\end{array}
$$

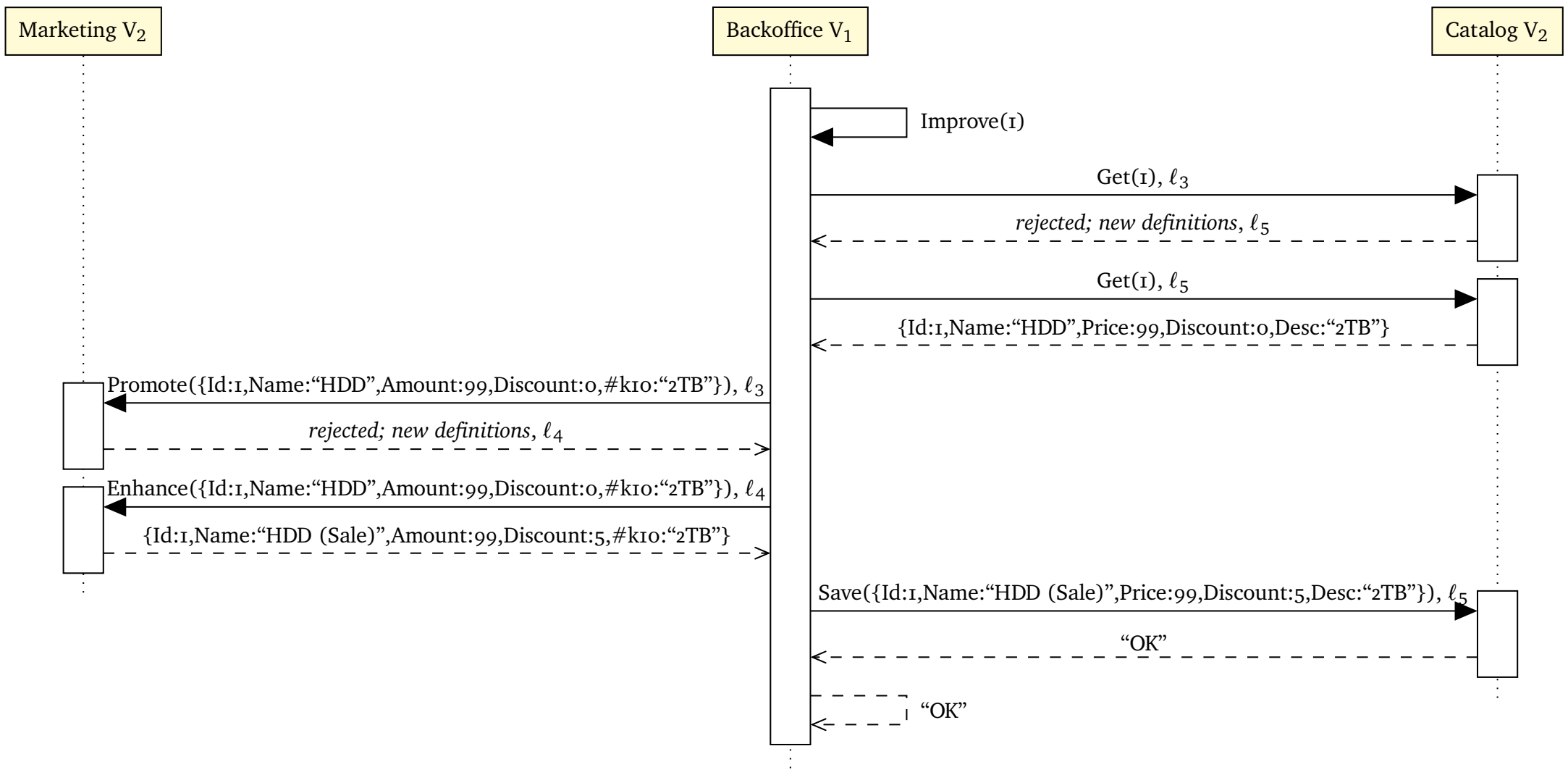

**Figure 4** Calling Backoffice.Improve after deploying versions $V_2$ of Catalog, Marketing.

**Figure 5** The language of expressions

modules $a$, functions $f$, types $n$, element keys $k$, record labels indexed by their keys $r_k$, deployment labels $\ell$, and identifiers $t, x$. We define a closed syntactic category $(\beta, \alpha, \delta)$ of base types to represent a typical data exchange format (cf. JSON), and from which we omit the trivial extension to lists

$$\beta, \alpha, \delta ::= \texttt{int} \mid \texttt{string} \mid \{\overline{r_k : \beta}\} \mid n_k$$

and define the type language $(\tau, \sigma)$ with base types and arrow types

$$\tau, \sigma ::= \beta \mid \tau \to \tau$$

We then adopt a simply typed lambda calculus with records as our core language. The language, defined in figure 5, includes function names ($f$), identifiers ($x$), lambda expressions ($\lambda x : \tau.e$), and function calls ($e(e)$). Record operations include a constructor ($\{\overline{r_k = e}\}$) and a destructor (selection $e.r$), and the ability to functionally modify a record field, by creating a modified copy ($e\,\{\overline{r_k = e}\}$). Expression $t?$ denotes the result of an expression execution on a thread $t$ (cf. a future or promise). Futures are here used to implement inter-service communication and can be used in both a synchronous or asynchronous communication style. The semantics we defined for our core language is call-by-value and implements synchronous communication.

Given the language of expressions, which may be instantiated in any mainstream programming language, we define a coordination language that captures the definition of modules and services in figure 6. A system ($U$) consists of a set of services. A service ($S$) is a runtime entity that has a module definition ($M$), a set of proxies to other services ($\overline{P}$), a deployment label ($\ell$), and a set of running threads ($\overline{T}$). A module is a static entity (code). It is defined by a name ($a$), a set of references to other modules ($\overline{R}$), and a set of type and function definitions ($\overline{D}$). Composition operator $S \parallel U$ defines a system with two separate parts, hosting disjoint sets of service/module names. We use $\overline{S}$ interchangeably to denote a system and a set of services. Each type and value definition in a module has a unique key ($k$). A module contains references with the signatures of all its consumed elements. Each module reference consists of the name of the producer module, and a set of type and value references ($\overline{X}$). Type and value references capture relevant signature information from the element they refer to, namely the element's producer module, name, key, and type definition.

Note that the types in all definitions and references are restricted to base types ($\beta$). On the other hand, expressions in general don't need this restriction, as noted by the use of $\tau$ in the parameter type of the abstraction expression ($\lambda x : \tau.e$). For the sake of simplicity, besides function types, we omit other unrestricted composite types from $\tau$ (e.g. record types of the form $\{\overline{r_k : \tau}\}$) and list types without any loss of generality.

Proxies of a service refer to another service implementing a given module. It contains a set of value proxies, referring to individual definitions in the producer service/module ($\mathsf{proxy}(a, \overline{Y}, \ell)$), or a proxy placeholder containing enough information to create an up-to-date proxy ($\mathsf{update}(a, \Phi, \ell)$). They store the name of the producer module and the expected deployment label of the producer service. In the case of value proxies, it carries the local name of the mapped function, and the corresponding name and type in the producer module (at the time the proxy was created). In the case of the placeholder, it carries a deployment label and a signature that is enough to produce new value proxies. Finally, each service hosts a set of threads, which have unique identifiers and associated expressions. Threads are silently executed by the system.

For readability, the examples presented in section 3 and section 4 use a mix of textual and graphical notations, where the textual notation is based on the syntax just defined. In listing 1 we illustrate how the system depicted in figure 3b can be defined using the syntax just presented. By comparing this code wit h the examples it should be clear how one representation can be mapped to the other.

To simplify the presentation of the semantic rules in the next sections, the declaration of a service $\mathsf{service}(M, \overline{P}, \ell, \overline{T})$ for module M $\triangleq$ $\mathsf{module}(a, \overline{R}, \overline{D})$ is going to be presented as $a\big(\,\overline{T}\,\big)$. This representation shows the set of threads $\overline{T}$ currently being executed by service $a$. Remaining service components can be accessed through a set of functions: $a_{\mathsf{refs}}$ denotes the references $\overline{R}$; $a_{\mathsf{defs}}$ denotes the definitions $\overline{D}$; $a_{\mathsf{proxies}}$ denotes the proxies $\overline{P}$; and $a_{\mathsf{label}}$ denotes the (unique) service label $\ell$. Components of M can be accessed by similar functions: $M_{\mathsf{name}}$ denotes the module name, $a$; $M_{\mathsf{refs}}$ denotes the references $\overline{R}$; and $M_{\mathsf{defs}}$ denotes the definitions $\overline{D}$.

$$
\begin{array}{rll}
U & ::= \emptyset \mid S \parallel U & \textit{(system)} \\
S & ::= \mathsf{service}(M, \overline{P}, \ell, \overline{T}) & \textit{(service)} \\
M & ::= \mathsf{module}(a, \overline{R}, \overline{D}) & \textit{(module)} \\
R & ::= \mathsf{ref}(a, \overline{X}) & \textit{(module ref)} \\
X & ::= \mathsf{typeR}(a, n, k, \beta) & \textit{(type reference)} \\
 & \mid\ \mathsf{valueR}(a, f, k, \beta \to \beta) & \textit{(value reference)} \\
D & ::= \langle k, n \rangle = \beta & \textit{(type definition)} \\
 & \mid\ \langle k, f \rangle : \beta \to \beta = e & \textit{(value definition)} \\
P & ::= \mathsf{proxy}(a, \overline{Y}, \ell) & \textit{(proxy)} \\
 & \mid\ \mathsf{update}(a, \Phi, \ell) & \textit{(outdated)} \\
Y & ::= \mathsf{valueP}(f, f, \beta \to \beta) & \textit{(value proxy)} \\
\Phi & ::= \overline{k : \tau} & \textit{(signature)} \\
T & ::= t \rhd e & \textit{(thread)}
\end{array}
$$

**Figure 6** The language of services.

## 6 Type System

We now define a type system that guarantees the absence of all communication errors. The typing relation for systems is defined with judgment $C; \Gamma \vdash S : P$, where $C$ and $P$ are mappings from keys to tuples describing services $k : \langle b, f, \tau, \ell \rangle$ and type definitions $k : \langle b, n, \tau, \ell \rangle$. Each key maps to an element in a service $b$, with local function name $f$ (or local type name $n$), type $\tau$, and deployment label $\ell$. Each system fragment is typed with relation to other services, described by $C$, and provides the elements described in $P$. We also use a global typing environment to describe threads running in services, $\Gamma = \overline{t : th(\tau)}$ where type $th(\tau)$ describes a thread that yields values of type $\tau$.

Compound systems are two mutually dependent fragments, typed together by rule

$$
\frac{C, P'; \Gamma \vdash S : P \qquad C, P; \Gamma \vdash S' : P'}{C; \Gamma \vdash S \parallel S' : P, P'}
$$

Notice that two parts of a system may use the endpoints of each other in a mutually recursive composition and that the endpoints provided by the composition $S \parallel S'$ are the combination of the two $(P, P')$. The provided services and types of service $S$, denoted by $P$, are used in the typing of service $S'$ and vice-versa.

The base case of the typing derivations on systems is the typing of single services, captured by the typing rule

$$
\frac{\begin{array}{c} a = M_{\mathsf{name}} \quad a \notin C_{\mathsf{name}} \quad C \vdash M : \overline{\langle n, k \rangle : \delta} \quad \overline{P} = \overline{\mathsf{proxy}(b, \overline{Y}, \ell')} \cdot \overline{\mathsf{update}(c, \Phi, \ell'')} \\ \overline{\overline{C \vdash^{\ell'}_{M} Y\ \mathsf{ok}}} \quad \overline{|M_{\mathsf{defs}}|, |M_{\mathsf{refs}}|; \Gamma \vdash e : \sigma} \quad \overline{\Gamma(t) = th(\sigma)} \end{array}}{C; \Gamma \vdash \mathsf{service}(M, \overline{P}, \ell, \overline{t : e}) : \overline{k : \langle a, n, \delta, \ell \rangle}}
$$

**Listing 1** Code of system in figure 3b.

```
modCatalogV2 ≜ module(Catalog, {},
      { <k1,Product> = { Id_k2: int, Name_k3: string, Price_k4: int, Discount_k5: int, Desc_k10: string }
        <k6,Get>: int → Product_k1 =λp:int . { Id_k2 = 1, Name_k3 = "HDD", Price_k4 = 99, Discount_k5 = 0, Desc_k10 = "2TB" }
        <k7,Save>: Product_k1 → string = λid:int . "OK"
      })

modMarketingV2 ≜ module(Marketing,
      { ref(Catalog, { typeR(Catalog, Product, k1, { Id_k2: int, Name_k3: string, Amount_k4: int, Discount_k5: int })})
      }
      { <k8, Enhance>: Product_k1 → Product_k1 = λp:Product_k1 . p { Name_k3 = p.Name + "Pro", Discount_k5 = 5 }
      })

modBackofficeV1 ≜ module(Backoffice,
      { ref(Catalog, { typeR(Catalog, Product, k1, { Id_k2: int, Name_k3: string, Amount_k4: int, Discount_k5: int }),
                       valueR(Catalog, Get, k6, int → Product_k1),
                       valueR(Catalog, Save, k7, Product_k1 → string)}),
        ref(Marketing, { valueR(Marketing, Facelift, k8, Product_k1 → Product_k1)})
      },
      { <k9,Improve>: int → string = λid:int . Save(Facelift(Get(id)))
      })

service(modCatalogV2, {}, ℓ5, {})

service(modMarketingV2, {}, ℓ4, {})

service(modBackofficeV1,
      { proxy(Catalog,
              { valueP(Get, Get, int → { Id_k2: int, Name_k3: string, Price_k4: int, Discount_k5: int, Desc_k10: string }),
                valueP(Save, Save, { Id_k2: int, Name_k3: string, Price_k4: int, Discount_k5: int, Desc_k10: string } → string)
              }, ℓ5),
        proxy(Marketing, { valueP(Facelift, Enhance, { Id_k2: int, Name_k3: string, Amount_k4: int, Discount_k5: int } →
                                  { Id_k2: int, Name_k3: string, Amount_k4: int, Discount_k5: int }

                       }, ℓ4)
      }, ℓ3, {})
```

where $C_{\mathsf{name}}$ denotes the set of the names of the modules in $C$. The rule encloses basic validations such as the uniqueness of the service/module name $a \notin C_{\mathsf{name}}$, the typing of the enclosed module $M$, its proxies $\overline{P}$ and expressions active in threads $\overline{t : e}$. Moreover we use the notation $|\langle n, k\rangle : e = \tau| \triangleq \overline{n : \tau}$, and $|\overline{\mathsf{valueR}(b, f, k, \tau)}| \triangleq \overline{f : \tau}$ to extract a typing environment from the definitions ($M_{\mathsf{defs}}$) and references ($M_{\mathsf{refs}}$) of the module, respectively. The typing of services is dependent on the typing relation on modules, given by the judgment $C \vdash M : P$. Again, $C$ describes the resources of other modules and $P$ describes the interface provided by module $M$. The typing of modules statically validates that the module references are in sync with the current knowledge about the system ($C$), and that the module definitions are well-typed with relation to its definitions and references. Within a service, the type signatures stored in proxies must match when the deployment label also matches the know producer's label ($\overline{\overline{C \vdash^{\ell'}_{M} Y \text{ ok}}}$) – notice that we have a list of list of individual proxies in each module ($\overline{\overline{Y}}$). The definition of a module is hence verified by the typing rule,

$$\frac{\begin{array}{c}\overline{X_b} = \overline{\mathsf{valueR}(b, f, k', \delta')} \quad C; \Sigma' \vdash_{\mu} \overline{X_b} \text{ ok} \quad \mu = \mathscr{U}(\overline{X_b}, \overline{D}) \quad \forall_b \\ \overline{D} = \overline{\langle n, k\rangle : \delta = e}, \overline{\langle m, k\rangle = \beta} \\ \Sigma' = \overline{\overline{m' = \beta'}} \quad \Sigma = \Sigma', \overline{m = \beta} \quad \Delta = \overline{\overline{f : \mathsf{expand}(\tau, \Sigma')}}, \overline{n : \mathsf{expand}(\delta, \Sigma)} \quad \Sigma;\ \Delta \vdash D_i : P_i \quad \forall_{i \in 1..|\overline{D}|}\end{array}}{C \vdash \mathsf{module}(a, \overline{\mathsf{ref}(b, \overline{X_b})}, \overline{D}) : \overline{P}}$$

One important invariant of the typing judgment for expressions is that all type names ($n_k$) in environments $\Delta$ and $\Gamma$ are always expanded (denoted by $\mathsf{expand}(.,.)$ in the rule above) with relation to the definitions in $\Sigma$.

We use the compact notation $\overline{\overline{X_b}}$ for sequences of sequences of items, in this case, of references, each collection belonging to a different service with name $b$. Notation $\mathscr{U}(\overline{X_b}, \overline{D})$ denotes the keys from service $b$ used in the current module, definitions $\overline{D}$.

**Definition 1** (Compatibility Relation). *We inductively define a compatibility relation on types, written $\tau \overset{\mu}{\rightsquigarrow} \sigma$, and read "$\tau$ evolves through compatible changes to $\sigma$ with a set of keys $\mu$".*

$$\begin{aligned} \beta \overset{\mu}{\rightsquigarrow} \beta &\triangleq \texttt{true} \\ \tau_1 \to \tau_2 \overset{\mu}{\rightsquigarrow} \tau'_1 \to \tau'_2 &\triangleq \tau'_1 \overset{\mu}{\rightsquigarrow} \tau_1 \wedge \tau_2 \overset{\mu}{\rightsquigarrow} \tau'_2 \\ \{\overline{r_{k_1} : \beta_1}\} \overset{\mu}{\rightsquigarrow} \{\overline{r'_{k_2} : \beta_2}\} &\triangleq \forall_{r_{k_1}:\beta_1}.k_1 \notin \mu \ \vee \exists_{r'_{k_1}:\beta_2}.\beta_1 \overset{\mu}{\rightsquigarrow} \beta_2 \end{aligned}$$

This relation is a unidirectional and reflexive relation. In the case of function types, it is covariant in the results and contravariant in the parameter types. Record types are related by the keys of the fields, not the names, which is the basic support for renaming of labels, and for the removal of unused fields ($k1 \notin \mu$). This means that the soundness of a system depends on the compatibility between the actual types in a producer service and the types stored in the references of consumer services. This compatibility relation establishes the minimum requirements to convert values from one type to another without loss of information.

Finally, the typing is defined by a judgment of the form $\Sigma; \Delta; \Gamma \vdash e : \tau$. In this definition we proceed with the expansion of named types whenever necessary. Use of named types allows for a greater control in the evolution of interfaces without loosing the semantics of values stored in some service or in transit in the system.

Notice that a system has an abstract signature (a type) that vouches for its soundness at runtime. The typing of a system is an invariant of the semantics for systems, given an initial state is obviously sound (e.g. the empty system). Nevertheless, the typing of systems implies the typing of the modules therein. This is a relation established by a compile-time procedure and is a precondition for the deploy operation along with other deploy-time verifications that we define at this point.

We next present the compatibility relation, a central piece in our system, that works like a preflight check for the deployment of modules. It is performed by the *global* *deployment manager* and is based only on the meta-information stored in that component, comprising the signatures of all modules already deployed in services and the signatures of the modules being deployed. We present some notation and intermediate results that support the module compatibility relation and the disconnected module relation, needed to validate the undeployment of services.

A signature $\Phi$ is defined as a mapping from (unique) keys to types, of the form $\overline{k:\tau}$. The signature of a module $M$, with name $m$, written $\Phi_m$, contains the mapping from keys to (fully expanded) types of all definitions in module $M$. We lift the definition for sets of modules, written $\Phi_{\overline{m}}$, by concatenating all module signatures $\overline{\Phi_m}$. A signature of a system $S$, written $\Phi_S$ is defined by the concatenation of the module signatures $\overline{\Phi_m}$ that are deployed in services inside the system. Given a module $M$ with name $m$, a producer signature, written $\rho_m$, is defined by the set of keys in all definitions of module $M$. The consumer signature of a module, written $\theta_m$, is defined by the set of keys in all references of module $M$. We also use the definition for sets of modules, $\theta_{\overline{m}}$. For a system $S$, we define its consumer signature, written $\theta_S$, as the set of keys in all references of the modules $M$ in $S$. Furthermore, we use the notation $S\backslash\overline{m}$ to denote the set of services in $S$ whose name is not in $\overline{m}$. We also use $\Phi_{S\backslash\overline{k}}$ to denote the set of key-type assignments in $\Phi_S$ whose key is not in $\overline{k}$. Given these auxiliary definitions and notation we next define the compatibility of a set of modules with relation to a system. Intuitively, the compatibility verification checks if all the items replaced by the modules in the system will be compatible to the ones effectively used by the remaining services, and if the requirements of the new modules are satisfied by the existing resources in the system. We show that this condition, that is only defined for well-typed sets of modules, is enough to ensure that the deployment of such a set of modules will not disrupt the operation of a well-formed system.

**Definition 2** (Module compatibility). *A set of modules $\overline{M}$, such that $C \vdash \overline{M} : P$, is compatible with a deployed system $S$, stated as $\mathscr{X}(S,\overline{M},C,P)$, by*

$$\mathscr{X}(S,\overline{M},C,P) \triangleq \begin{cases} \forall k \in \rho_{\overline{m}}.\ k \in \theta_{S\backslash\overline{m}} & \Longrightarrow P(k) = \langle b,n,\delta,\ell\rangle \\ & \quad \wedge\ \Phi_S(k) \rightsquigarrow_{\mu_{S\backslash\overline{m}}} \delta \\ \forall k \in \theta_{\overline{m}}.\ C(k) = \langle b,n,\delta,\ell\rangle & \Longrightarrow \delta \rightsquigarrow_{\mu_{\overline{m}}} \Phi_S(k) \end{cases}$$

The system after the deployment of modules $\overline{M}$, with names $\overline{m}$, has the signature $\Phi_{S'} = \Phi_S\backslash\rho_{\overline{m}} \cup \{k : \tau \mid k \in \rho_{\overline{m}} \wedge P(k) = \tau\}$.

Also, a set of services can only be undeployed without disrupting the system if they represent a disconnected part of the system, as in the following definition.

**Definition 3** (Disconnected Systems). *We say that system $S'$ is disconnected from system $S$, written $S\#S'$, if $\forall_{k\in\rho_{S'}},\ k \notin \theta_S$.*

Typing defines invariants that are preserved by the semantics presented next.

## 7 Deploy semantics

The operational semantics is defined using a labelled transition system that describes how a given ecosystem of services evolves. The transitions capture internal computations within a service, interactions between services, and service evolution steps. We design the operational semantics to not check the safety of deployment or undeployment operations. We model a regular cloud infrastructure that simply loads code without further checks. The operational semantics is twofold. It accounts for the execution of services, by evaluating active expressions within the services and by implementing deploy and undeploy operations. The bootstrap of the execution of expressions is represented by a special transition that starts the execution of an expression within a service.

We extend our syntax of expressions, defined in figure 5, such that record values can refer to both known and unknown fields ($\{\overline{r_k = v}, \overline{\#k = v}\}$). Unknown fields are identified by (unique) keys ($\#k$). This capability allows us to preserve data when it is exchanged between services that have a different (but compatible) definition of a type. As an example, consider two definitions of type Product that we've seen in our running example (figure 2). The first used by version 2 of Catalog and the second used by Marketing and Backoffice

$$\{\cdots, \mathit{Price}_{k4} : \mathtt{int}, \mathit{Discount}_{k5} : \mathtt{int}, \mathit{Desc}_{k10} : \mathtt{string}\}$$
$$\{\cdots, \mathit{Amount}_{k4} : \mathtt{int}, \mathit{Discount}_{k5} : \mathtt{int}\}$$

Here, one field was renamed (blue-shaded), and a new field was added (red-shaded). Consider the record values below, whose structures are their runtime representations:

$$\{\cdots, \mathit{Price}_{k4} = 99, \mathit{Discount}_{k5} = 0, \mathit{Desc}_{k10} = \text{“2TB”}\}$$
$$\{\cdots, \mathit{Amount}_{k4} = 99, \mathit{Discount}_{k5} = 0, \#k10 = \text{“2TB”}\}$$

The conversion ensures the data of new field Desc is preserved when interacting with services that are still using an older version of the type Product. Moreover, field name changes are detected and addressed.

We present in figure 7 the relevant operational semantics rules of our language. For instance, rules (Invoke), (Reject), and (GenProxy) describe the invocations of remote services. The evaluation of remote calls establish an interaction between a consumer and a producer service. We model these interactions by means of locating threads on services and having a runtime expression that signals that a thread is waiting for the result of a remote thread $t$ ($t?$).

Rule (Invoke) defines the case where some service $a$ successfully invokes a remote function on some service $b$. The second premise ensures that the proxies for service $b$ are up-to-date with the latest deployment of $b$, as label $\ell$ matches the current label for that service. This invocation creates a new thread in service $b$ to execute function $f$, whose name in $b$ is $f'$. Notice the conversion of types at the input and output of the function. In practice, both the input conversion (call to $\mathsf{convert}_{\tau}^{\tau'}$) and the output conversion (call to $\mathsf{convert}_{\sigma'}^{\sigma}$) are computed at the consumer service.

$$\dfrac{\begin{array}{c} a_{\text{proxies}} = \text{proxy}(b, \text{valueP}(f, f', \tau' \to \sigma') \cdot \overline{Y}, \ell) \cdot \overline{P_a} \\ b_{\text{label}} = \ell \quad a_{\text{refs}}(b) = \text{valueR}(b, f, k, \tau \to \sigma) \cdot \overline{X} \quad s \text{ fresh} \end{array}}{\begin{array}{c} a\big( t \rhd \mathscr{C}[f(v)] \mid \overline{T_a} \big) \,\Big\|\, b\big( \overline{T_b} \big) \longrightarrow \\ a\big( t \rhd \mathscr{C}[\text{convert}^{\sigma}_{\sigma'}(s?)] \mid \overline{T_a} \big) \,\Big\|\, b\big( s \rhd f'(\text{convert}^{\tau'}_{\tau}(v)) \mid \overline{T_b} \big) \end{array}} \ (\textsc{Invoke})$$

$$\dfrac{}{a\big( t \rhd \mathscr{C}[s?] \mid \overline{T_a} \big) \,\Big\|\, b\big( s \rhd v \mid \overline{T_b} \big) \longrightarrow a\big( t \rhd \mathscr{C}[v] \mid \overline{T_a} \big) \,\Big\|\, b\big( \overline{T_b} \big)} \ (\textsc{Resolve})$$

$$\dfrac{\begin{array}{c} a_{\text{refs}}(b) = \text{valueR}(b, f, k, \tau) \cdot \overline{X} \\ \mathit{labelOf}(a_{\text{proxies}}(b)) = \ell \quad b_{\text{label}} = \ell' \quad \ell' \neq \ell \quad b_{\text{sig}} = \Phi \\ a' = a[\text{proxies} \mapsto \text{update}(b, \Phi, \ell') \cdot \overline{P_a}] \end{array}}{a\big(t \rhd \mathscr{C}[f(v)] \mid \overline{T_a}\big) \Big\| b\big(\overline{T_b}\big) \longrightarrow a'\big(t \rhd \mathscr{C}[f(v)] \mid \overline{T_a}\big) \Big\| b\big(\overline{T_b}\big)} \ (\textsc{Reject})$$

$$\dfrac{\begin{array}{c} \overline{\ell} \text{ fresh} \quad |\overline{\ell}| = |\overline{M}| \quad \overline{a_c} \subseteq \overline{a} \quad \forall_{i \in 1..|\overline{M}|} \quad M_i = \text{module}(a_i, \overline{R}, \overline{D}) \\ a_i = \text{service}(M_i, \{\ \text{proxy}(b, \emptyset, \ell_i) \mid \text{ref}(b, _) \in \overline{R}\ \}, \ell_i, \emptyset) \end{array}}{S \,\Big\|\, \overline{c(\emptyset)} \xrightarrow{\text{deploy}\,\overline{M}} S \,\Big\|\, \overline{a(\emptyset)}} \ (\textsc{Deploy})$$

$$\dfrac{\begin{array}{c} b_{\text{sig}} = \Phi \quad a_{\text{proxies}} = \text{update}(b, \Phi, \ell) \cdot \overline{P_a} \\ \overline{Y} = \mathscr{P}(b, a_{\text{refs}}, \Phi) \\ a' = a[\text{proxies} \mapsto \text{proxy}(b, \overline{Y}, \ell) \cdot \overline{P_a}] \end{array}}{a\big(\overline{T}\big) \longrightarrow a'\big(\overline{T}\big)} \ (\textsc{GenProxy})$$

$$\dfrac{s \text{ fresh}}{a\big(\overline{T}\big) \xrightarrow{a:\text{start}(e)} a\big(s \rhd e \mid \overline{T}\big)} \ (\textsc{Start})$$

**Figure 7** A selection of operational semantics rules.

$$\mathsf{convert}_{\tau}^{\tau}(x) \triangleq x$$

$$\mathsf{convert}_{\tau\to\sigma}^{\tau'\to\sigma'}(f) \triangleq \lambda x : \tau'.\mathsf{convert}_{\sigma}^{\sigma'}(f(\mathsf{convert}_{\tau'}^{\tau}(x)))$$

$$\mathsf{convert}_{\{\overline{r_k:\alpha}\}}^{\{\overline{r'_{k'}:\beta}\}}(v) \triangleq \begin{cases} r'_{k'} = v.\#k & k' \in \mathsf{keys}(\{\overline{r'_{k'}:\beta}\}) \wedge \\ & k \in \mathsf{keys}(v) \ \wedge \mathsf{typeof}(v.\#k) = \beta \\ r'_{k'} = \mathsf{convert}_{\mathsf{typeof}(v.\#k)}^{\beta}(v.\#k) & k' \in \mathsf{keys}(\{\overline{r'_{k'}:\beta}\}) \wedge \\ & k \in \mathsf{keys}(v) \ \wedge \mathsf{typeof}(v.\#k) \neq \beta \\ r'_{k'} = \mathsf{default}(\beta) & k' \in \mathsf{keys}(\{\overline{r'_{k'}:\beta}\}) \wedge \\ & k \notin \mathsf{keys}(v) \\ \#k = v.\#k & k' \notin \mathsf{keys}(\{\overline{r'_{k'}:\beta}\}) \wedge \\ & k \in \mathsf{keys}(\{\overline{r_k:\alpha}\}) \end{cases}$$

**Figure 8** Function $\mathsf{convert}_{\tau}^{\sigma}$ that adapts values of type $\tau$ to type $\sigma$.

Rule (Resolve) defines the termination of a function call started by (Invoke). Thread $s$ in service $b$ terminates and returns value $v$, while thread $t$ resumes execution.

Rule (Reject) defines the case where an invocation of function $f$ of remote service $b$ is rejected, because $b$ has evolved (has been redeployed) since its proxies in $a$ where generated. This mismatch is detected by the mismatch between the deployment label in the proxy ($\ell$) and the producer service label ($\ell'$). This causes the consumer module $a$ to initiate the generation of proxies for module $b$ according to its latest signature ($b_{\mathsf{sig}}$). We define an auxiliary function *labelOf* that returns the label of the proxies for a given remove service. We also define the signature of a service $b_{\mathsf{sig}}$ as the collection of all its function's signatures.

Whenever a proxy is outdated, it is replaced by a token ($\mathsf{update}(b, \Phi, \ell')$) containing the necessary signature to produce the proxy code, in our case emulated by the convert function. If a proxy is under update and the target signature changes, the stored signature is updated, guaranteeing that the generated proxy is correct.

Rule (GenProxy) shows the generation of proxies for the references in $a$ to the producer service $b$. The generation function $\mathscr{P}$ generates the proxies based on the updated information on service $b$. For each reference from $a$ to $b$, $\mathscr{P}$ registers: the local reference name; the corresponding name in the producer module $b$; and its type.

Rule (Deploy) expresses the deployment of a set of modules. Some of these modules are being redeployed after being redefined ($\overline{c}$), while the remaining ones are new deployments. A module can only be stopped if it is quiescent, i.e. it has no threads running ($c(\emptyset)$). The proxy for each reference in a new service $a$ is empty and its assigned the corresponding service label ($\mathsf{proxy}(b, \emptyset, \ell_i)$). When the first interaction with a remote service occurs, the use of fresh labels ensures that there is a mismatch with the producer label, causing an update.

The initial configuration is an empty system, and systems are defined by sequences of deployments. Rule (Start) bootstraps the execution of expressions.

We show in figure 8 the definition for the adapter function used in rule (Invoke) to convert values between two compatible types. The first case of the definition is the identity function. The second case converts values of function type and recursively converts the argument (contravariantly) and return types. The following cases deal with record values, and depict situations where unknown fields can exist both in the origin and the target type. In the first three cases for records all fields are known to the target type. In the first and second cases the field's key is also known in the source value (but not necessarily in the source type), the former does not require any conversion while in the later the field type has to be converted. In the third case, the field key does not exist in the source value, therefore it is initialised with a default value. The last case for records, deals with field keys of the source type that do not exist in the target type. Then, these fields are preserved in the target as unknown fields, thus ensuring that no information is lost by the type conversion. Note that the primitive operator $v.\#k$ is used both to access known fields ($r_k$) and unknown fields ($\#k$) as the keys are unique inside a record.

## 8 Technical Results

We next present the results of both our model and system architecture, specifically we show that the deployment of new modules preserves the well-formedness of the microservice system (theorem 1). A corollary of this result is that starting from an empty system, and only performing verified deployment actions, we never reach a system with incompatible interfaces between microservices.

We incrementally present partial results that lead to the final theorem. We separately prove the preservation of types for expressions, internal system reductions (computations) and reductions caused by actions like deploy, undeploy and spawning of new threads. In this section, we sketch the proofs of the lemmas and theorems below.

Any well-typed expression is guaranteed to reduce in the context of a set of definitions ($\overline{D}$) to an expression with the same type and therefore yield results of that type.

**Lemma 1** (Type preservation for expressions). *If* $\Sigma; \Delta \vdash e : \tau$*, and* $\Sigma; \Delta \vdash \overline{D} : \overline{P}$*, and* $\overline{D} \vdash e \longrightarrow e'$ *then* $\Sigma; \Delta \vdash e' : \tau$.

*Proof.* By induction on the size of the typing derivation and by case analysis on the last reduction rule applied. □

In this next step, given that well-typed expressions preserve typing (lemma 1), we prove that all computations, that don't change the structure of the system – inside services or service invocations, performed by the services keep the well-formedness of the systems.

**Lemma 2** (Preservation on computation). *If* $C; \Gamma \vdash S : P$*, and* $S \longrightarrow S'$ *then* $C; \Gamma' \vdash S' : P$ *for some* $\Gamma'$.

*Proof.* By induction on the size of the reduction derivation and by case analysis on the last reduction rule applied. For all cases, rules (T-SERVICE) and (T-SYSTEM) are used to breakdown $S$ and reassemble $S'$. In the case for rule (INVOKE) we leverage knowledge that calls to definitions have arguments of base types ($\beta$) to conclude the argument $v$ can be typed in an empty environment, a stepping stone to prove that the value can effectively be transported to the producer service and typed correctly there. For rule (GEN-PROXY), we explore the possibility that the deployment label ($\ell$) of the proxy may or may not be up-to-date with the timestamp of the producer service, proceeding the type derivation with (T-PROXY) or (T-OUTDATED) respectively. ☐

We now prove that all new expressions being spawned that are well-typed in the context of a module maintain the well-formedness of the system and therefore do not introduce any disruption.

**Lemma 3** (Preservation on start expression). *If $C;\Gamma \vdash S : P$ and module*

$$M_b = \mathit{module}(b, \overline{R}, (\overline{\langle n,k\rangle : \delta = e}, \overline{\langle n',k'\rangle = \delta'}))$$

*with references*

$$\overline{R} = \overline{\mathit{ref}(a, (\overline{\mathit{valueR}(b,m,k',\sigma)}, \overline{\mathit{typeR}(a,m',k',\sigma')}))},$$

*and $S \xrightarrow{b:start(e)} S'$, and expression $\overline{\overline{m' : \sigma'}}, \overline{n' : \delta'}; \overline{\overline{m : \sigma}}, \overline{n : \delta} \vdash e : \tau$, then there is $\Gamma'$ such that $C;\Gamma' \vdash S' : P$.*

*Proof.* By inspection of the typing and reduction rules. The only reduction rule applicable in this case is rule (START). By inversion of typing derivation we know that all expressions in threads $\overline{T_b}$ in a service $b(\,\overline{T_b}\,)$ are well-typed, $\Gamma, |M_{\mathsf{defs}}|, |M_{\mathsf{refs}}| \vdash e : \sigma$. The new expression $e$ is well-typed as hypothesis $(\overline{\overline{m' : \sigma'}}, \overline{n' : \delta'}; \overline{\overline{m : \sigma}}, \overline{n : \delta} \vdash e : \tau)$ and we have $|M_{\mathsf{defs}}| = \overline{n : \delta}$ and $|M_{\mathsf{refs}}| = \overline{\overline{m : \sigma}}$. Hence, by applying rule (T-SERVICE) we conclude that the new service $b(\,s : e \mid \overline{T_b}\,)$ is also well typed with $\Gamma' = \Gamma, s : th(\tau)$. Hence, the system is also well-typed by reconstructing the derivation with rule (T-SYSTEM). ☐

We now look into the deploy operation and prove that all new services being deployed are well-typed and in the context of the existing system maintain its well-formedness and therefore do not introduce any disruption. Notice the combination of static type checking of modules being deployed ($\overline{M}$) with the runtime validation of their interface specification with the actual system metadata by function $\mathscr{X}$ (performed by the global deployment manager of the system).

**Lemma 4** (Preservation on Deployment). *If $S \xrightarrow{deploy\,\overline{M}} S'$, and $C;\Gamma \vdash S : P, P'$ and $\forall_{i \in 1..|\overline{M}|}.\ C, P, \overline{P''} \vdash M_i : P''_i$, and $\mathscr{X}(S, \overline{M}, C, \overline{P''})$, then $C;\Gamma \vdash S' : P, \overline{P''}$.*

*Proof.* By inspection of the reduction derivation and by analysis of the typing derivation. The only rule available for deploy derivations is Rule (DEPLOY), where the application of (T-SERVICE) to the new services is trivial as there are no proxies or threads present, and the hypothesis states that the modules are well-typed. Given a well-typed set of services from the modules given, we have to show that the requirements of the remaining system are satisfied by the newly deployed services and vice-versa. This is achieved by lifting the notion of compatibility from types to environments and proving a property of weakening based on compatibility. □

**Definition 4** (Disjoint environments). *Two environments, $C$ and $P$, are disjoint, written $C\#P$, if their domains are disjoint and the set of module names are disjoint.*

The same reasoning needs to be done for the undeployment of a set of microservices.

**Lemma 5** (Preservation on Undeployment). *If $S \xrightarrow{\mathsf{undeploy}\,\overline{c}} S'$, and $S\#\overline{c}$, with $\overline{c}_{defs} = \overline{\langle n,k\rangle : \delta = e}, \overline{\langle n,k\rangle = \delta}$ and $C \vdash S : P, \overline{k : \langle b,n,\delta,\ell\rangle}$ then $C;\Gamma \vdash S' : P$.*

*Proof.* By inspection of the reduction and typing derivation with only (Undeploy) as the last possible reduction rule. □

The global type preservation result gathers the previous lemmas in a single result.

**Theorem 1** (Type preservation). *For all systems $S$, and type assertions $C,\Gamma,P$, if the starting system $S$ is well-formed ($C;\Gamma \vdash S : P$) then*

1. *if $S$ evolves internally ($S \longrightarrow S'$), then $S'$ is well-formed ($C;\Gamma \vdash S' : P$), and*
2. *if an expression $e$ is well-typed, written $\overline{\overline{m:\sigma}, n:\delta} \vdash e : \tau$, with relation to some module $M_b = \mathsf{module}(b, \mathsf{ref}(a, \overline{\mathsf{valueR}(b,m,k',\sigma)}), \overline{\langle n,k\rangle : \delta = e})$, and it is started in service $b$ ($S \xrightarrow{b:\mathsf{start}(e)} S'$), then $S'$ is well-formed ($C;\Gamma' \vdash S' : P$) for some $\Gamma'$, and*
3. *if a set of well-typed modules $\overline{M}$ ($\forall_{i\in 1..|\overline{M}|}.\ C,P_0,\overline{P''} \vdash M_i : P''_i$), that are compatible with the system ($\mathcal{X}(S,\overline{M},C,\overline{P''})$) is deployed $S \xrightarrow{\mathsf{deploy}\,\overline{M}} S'$ with $P = P_0, P'$, then the target system is well-formed ($C;\Gamma \vdash S' : P_0,\overline{P''}$), and finally*
4. *if a set of modules $\overline{c}$, disjoint from the system $S$ ($S\#\overline{c}$) is undeployed ($S \xrightarrow{\mathsf{undeploy}\,\overline{c}} S'$), then the target system is well-formed, and the undeployed services are no longer available $C;\Gamma \vdash S' : P\backslash\overline{k : \langle b,n,\delta,\ell\rangle}$, with $\overline{c}_{defs} = \overline{\langle n,k\rangle : \delta = e}$.*

*Proof.* Each item of the theorem is proven by lemmas 2, 3, 4, and 5. □

The preservation of types guarantees that the system always evolves in a principled manner, abiding by type compatibility between the different producer and consumer services. But the dynamically generated proxies and unknown fields, presented in section 7, also guarantee that no values are lost even when some services do not agree on the same type definition, effectively ensuring a preservation of values across calls. A preliminary result reflecting value preservation by the $\mathsf{convert}^{\sigma}_{\tau}$ function used in proxies is presented below.

**Lemma 6** (Compatibility implies no loss of values). *If types $\tau,\tau'$ are compatible $\tau \rightsquigarrow \tau'$ and there is a value $v'$ of type $\tau'$, then we have that $\mathsf{convert}^{\tau'}_{\tau}(\mathsf{convert}^{\tau}_{\tau'}(v')) = v''$ where either $v'' = v$ if no unknown fields exist in both $\tau$ and $\tau'$, or otherwise we know $v'' = \{\overline{r'_{k'} = v'.\#k'}, \overline{\#k = \mathsf{default}(\alpha)}\}$.*

*Proof.* By induction on the structure of the type $\tau'$. The case for record types requires exercising the combinations between known and unknown fields of both $\tau$ and $\tau'$, with the key aspect being that the inner conversion ($\mathsf{convert}^{\tau}_{\tau'}$) preserves all the fields from $\tau'$ that are not present in $\tau$ as unknown values, allowing the outer conversion ($\mathsf{convert}^{\tau'}_{\tau}$) to read them back. □

Another essential property that completes type safety in this setting is the progress of well-typed systems. In this case, it includes that a well-typed system should be composed only of terminated threads (whose expression is a value), and that for all threads waiting for a result there is a waiting thread expression $t?$ introduced by an external call. From inspecting the reduction rules we notice that threads are introduced by rule (Invoke) and eliminated by rule (Resolve). We can also observe that an expression being started in a service is not typed in an environment containing thread values, and thus cannot include waiting thread expressions. Also, threads cannot be passed around as first-class values. These arguments contribute for a well-founded, and paired, order of threads and corresponding waiting expressions. The proof technique necessary for such property should account for the role a thread has in system (caller or callee), discipline the linearity of thread values and detect deadlocks caused by entanglement of threads. Many works in the literature deal with this problem in similar situations [8, 9, 11] that should be applicable here.

## 9 Experimental Results

Our research is motivated by data from real, large-scale, applications created with the OutSystems platform. The evaluation of our model presented in this section is twofold. In a first assessment, we gather data from real applications and analyse the traces of past deployments (5 years) under our evolution model to measure its impact. In a second phase, we compare the actual evolution of an OutSystems application before and after the adoption of our compatibility model between module references.

**Analysis of evolution traces** We collected data about the evolution of three large software factories (collections of applications from third-party partner companies) in a development period of 5 years, each one containing more than 1000 modules. Our measurements aggregate changes, by kind and frequency, of module signatures observable in the deployment of new modules. Our raw dataset consists of 8889 production deployments, with a total of 23986 signature changes. The data was collected by querying the history of deployments in the meta-data database of the OutSystems platform, allowing us to validate our model and analyse its impact with real data. Note that this data refers to applications developed using a version of the OutSystems platform that did not compile to microservice architectures. Nevertheless, module references used in those applications have a direct correspondence to REST APIs and provide a reliable measure for the impact of our approach.

Table 1 presents the number of occurrences of signature changes aggregated by kind and divided by software factory. Our data set contains diverse large-scale enterprise

**Table 1** Occurrences of signature changes grouped by kind.

| Kind | Occurrences of Changes | | | | | | | |
|---|---|---|---|---|---|---|---|---|
| | Factory 1 | | Factory 2 | | Factory 3 | | Total | |
| | # | % | # | % | # | % | # | % |
| New Optional Field* | 7345 | 50.63% | 2354 | 26.61% | 301 | 47.55% | 10 000 | 41.69% |
| Change Field Type | 2204 | 15.19% | 2629 | 29.72% | 32 | 5.06% | 4865 | 20.28% |
| Remove Field | 1625 | 11.20% | 2507 | 28.34% | 70 | 11.06% | 4202 | 17.52% |
| New Mandatory Field | 1725 | 11.89% | 586 | 6.62% | 108 | 17.06% | 2419 | 10.09% |
| Rename Field* | 1156 | 7.97% | 620 | 7.01% | 45 | 7.11% | 1821 | 7.59% |
| Reorder Fields* | 320 | 2.21% | 96 | 1.09% | 34 | 5.37% | 450 | 1.88% |
| Change to Optional* | 99 | 0.68% | 43 | 0.49% | 11 | 1.74% | 153 | 0.64% |
| Change to Mandatory | 33 | 0.23% | 11 | 0.12% | 32 | 5.06% | 76 | 0.32% |

* Changes that are compatible in our model and in version 11 of the OutSystems platform.

applications from different domains (banking, insurance, and IT management), and shows some commonalities in the distribution of the kind of changes observed, but not a direct correspondence, as it can be observed from the relative frequency of changes in table 1. The correlation also exists when we observe the number of deployments where those changes occurred, shown in table 2. We can also observe that there is no direct correspondence between the ordering of changes related to frequency of deployments. We establish our baseline by reporting on changes that are deemed incompatible in version 10 of the OutSystems platform, i.e. currently for these kind of changes all consumer modules need to be explicitly adapted and new versions need to be deployed in order to avoid runtime errors. Note also that said changes are also incompatible, i.e. cause runtime errors, in any ad-hoc microservice implementations unless explicitly foreseen by the developers. In contrast, our model considers some of these changes as being compatible - the rows marked with an asterisk in table 1 correspond to compatible changes according to our model. The most frequent change that we are able to deem compatible is the addition of an optional field to a datatype being used in an interface. That alone accounts for an average of 41.7% of the changes introduced. The renaming of fields, reordering of fields in datatypes, and the changes from mandatory to optional are also changes that our model is capable of automatically and dynamically adapt and avoid the explicit correction and subsequent compilation of client modules' code. When combined, these changes account for 12424 signature changes, or 51.8% of all the changes in our data set. Other changes like changing a field's datatype, and addition of a mandatory field are not covered by our model. Our model also considers removal of unused fields, but since table 1 does not provide specific data about the usage of fields, we pessimistically assume that all removals are unsafe and not automatically adaptable.

Our data does not discriminate individual deployments that *only* contain compatible changes. Our data accounts only for the number of times a certain kind of change was observed in a deployment, and the number of separate deployments where they were observed (table 2). We proceed with the worst-case estimate for the reciprocal scenario – the number of deployments that contain *at least* one incompatible change – is obtained by assuming that all incompatible changes in our data occurred in separate deployments. For instance, in factory 1, the worst case scenario is when all kinds of changes in datatypes, removal of fields, new mandatory fields, and changes to the mandatory property of fields occur in separate deployments. That means that the deployment with breaking changes counting 2426 (723 + 877 + 798 + 28) in a total of 3759 deployments. That means that at least 35 % of the breaking deployments are automatically adapted. Table 3 presents the reduction on the number of incompatible signature changes after the adoption of our model. Results are better for the other two factories with reductions on the number of breaking deployments above 72 %. We conclude that an aggregated average of at least 57% of the analysed deployments are classified as safe under our model, and do not require the explicit adaptation, compilation, and deployment of new versions of any consumer modules. Hence, such deployments would be seamless to the users and cause the minimum amount of downtime and synchronization in a microservice architecture.

**Table 2** Deployments grouped by kind of signature changes.

| Kind | # Deployments | | | |
|---|---|---|---|---|
| | Factory 1 | Factory 2 | Factory 3 | Total |
| New Optional Field | 2192 | 1132 | 115 | 3439 |
| Change Field Type | 723 | 449 | 22 | 1194 |
| Remove Field | 877 | 493 | 45 | 1415 |
| New Mandatory Field | 798 | 357 | 38 | 1193 |
| Rename Field | 546 | 353 | 36 | 935 |
| Reorder Fields | 182 | 54 | 16 | 252 |
| Change to Optional | 80 | 43 | 4 | 127 |
| Change to Mandatory | 28 | 6 | 22 | 56 |

**Table 3** Safe and incompatible deployments analysed per factory.

| Source Factory | # Changes | Deployments | | | |
|---|---|---|---|---|---|
| | | Total | Broken | Safe | |
| | | # | # | # | % |
| Factory 1 | 14 507 | 3759 | 2426 | 1333 | 35.46% |
| Factory 2 | 8846 | 4659 | 1305 | 3354 | 71.99% |
| Factory 3 | 633 | 471 | 105 | 366 | 77.71 % |
| Total | 23 986 | 8889 | 3836 | 5053 | 56.85% |

**Compared analysis of signature compatibility** In a second assessment of our model, we evaluate the incremental evolution of a smaller monolithic application with approximately 450 modules, to a more loosely-coupled architecture where references are implemented by services and the compiler uses a partial implementation of our compatibility relation to determine what modules need adaptation and recompiling. This application is a development lifecycle management system that provides automation and supports a continuous integration environment inside OutSystems. The transition was enabled by upgrading the system to version 11 of the OutSystems platform, which supports service-based references and API compatibility analysis. All compatible changes enumerated in table 1 are used in version 11 to determine if a deployment has a breaking change requiring the intervention of a developer. Also, following our model, the implementation in version 11 includes the removal of fields based on actual usage, covered in our model, but not measured in the first analysis.

**Table 4** Modules automatically adapted before and after the compatibility model.

| | Deployments | Modules with changes | | Modules adapted | |
|---|---|---|---|---|---|
| | # | # | by deploy | # | by deploy |
| version 10 | 714 | 1679 | 2.4 | 7669 | 10.7 |
| version 11 | 559 | 1177 | 2.1 | 2423 | 4.3 |

We measured 714 production deployments during a period of 10 months of a monolithic style development, and observed 559 production deployments during the service-based phase also during 10 months. The results are shown in table 4, showing the number of modules that were explicitly changed by developers, and the number of modules that needed to be adapted as a consequence of the changes introduced. The average number of modules modified on each deployment reduced from 2.4 to 2.1, a slight descent that can be associated to the compatibility relation not demanding changes in consumer modules. The modules that required adaptation and recompilation per deployment was substantially reduced from an average value of 10.7 to 4.3 modules per deployment (59 % reduction). This result is not directly associated with the compatibility relation, but mostly a consequence of the adoption of a new criteria for detecting signature changes, similar to what happens in our model when the deployment label does not change. In version 11, module changes that do not modify the signature of programming elements (e.g. entities and functions) do not require consumer modules to be adapted. All remaining module adaptations needed in version 11 (4.3 modules per deployment) are identified as non-breaking in our model and are covered by the dynamic proxy mechanism presented in this paper. Such mechanism is not yet implemented in the platform, but the results presented here are already very promising.

In summary, we conducted two experiments. The first one uses aggregated data from past development histories evaluating them in the present model, and another that observed the actual evolution of an application using a version of the OutSystems platform that implements the compatibility relation presented in this paper. Our experiments point to a significant impact of our approach. However, the analysis can be refined to transform our worst-case results to a more fine-grained analysis with data that depicts each deployment individually and includes the actual combination of different kinds of changes on each deployment. The second experiment is an indirect comparison and measure of impact, but using the actual implementation of the compatibility model, although not yet in the realm of microservices.

## 10 Related Work

Service-oriented architectures have been stress-tested in past years, as systems and organisations adapt to make the most out of cloud technologies and deal with ever increasing amounts of data. Uber [21] and Netflix [26] are prominent examples of development of robust distributed systems designed with microservice architectures [13, 29, 35]. Software engineering guidelines assume a scenario of frequent service implementation changes and few interface or contract modifications [18], as breaking the contracts requires synchronization with consumers and may lead to runtime errors and unexpected behaviours. So, common designs focus on fault tolerant services that are robust to changes and failures on services they depend on, and ensure graceful degradation [15, 20, 36]. In contrast, our model focuses on reducing the types of changes that effectively breach the expected contracts.

Frameworks as Protobuf, Thrift, and Avro support data schema evolution [24, 25], but their support is more restricted and without the guarantees given by our approach. For instance, in Protobuf and Thrift new fields can be added. However, these fields must be optional to ensure newer service versions can read data written by older service versions. In Avro it is only possible to add or remove fields with a default value.

Breaking contracts is a concern that is already present in early studies about the evolution of library code [12]. In a loosely-coupled scenario, contract compliance through dynamic monitoring of semantic rules is a key, but error prone aspect. Approaches like [33] propose the use of API call logs to capture runtime dependencies between different services to predict the impact of deploying a new version of a service. Versioning of interfaces is adopted as a last resort, supported by conventions and dedicated methods [22, 23], as it adds an extra layer of complexity to systems developed by different teams. In contrast, our approach targets a setting where service interfaces and dependencies are statically-known by a lifecycle management system. The system leverages that information to assess: the impact of a deployment in all running consumer and producer services; and, the feasibility of generating compatible proxies. The need for versioning is eliminated, as the running services support automatic adaptation of interfaces for all compatible changes, effectively promoting a single-endpoint strategy.

In the context of service-oriented architectures (SOA) there has been extensive research on the adaptability of services [3, 4, 6, 7, 14, 27]. There are similarities between SOA and microservices, as both follow a separation of concerns approach based on services. These two architectures diverge on some relevant features, in particular on the fact the SOA services are tightly-coupled while microservices are loosely-coupled. As a consequence, adaptability in SOA is not focused on evolution but on the integration of microservices. These works [3, 4, 7, 14, 27] on SOA address and develop adapters for behavioural mismatches among business processes. The concern is mainly on adapting workflows and most assume that data adaptation is already defined by developers. Standard service binding technologies (cf. SOAP) use description languages like WSDL [1] to specify contracts, allowing them to check for compatibility of services on every remote call, and issue runtime exceptions when some mismatch happens. Schema registries and service brokers are also used to centralize and authorise changes in the service architecture. Schemas can then be checked based on version identifiers and compatibility relations just like in our case [22].

Specification and documentation technologies (cf. OpenAPI) provide a foundation for tools that check contracts between services [32], even at the semantic level [37]. Also, formal approaches applying behavioural types and static type checking ensure the soundness of service interfaces [9], even in a multiparty setting. None of the above accounts for evolution and adaptation of service contracts. We aim essentially, similar to [10], at tackling scenarios (and conditions) of evolution that allows one to replace one service and keep the whole system running.

## 11 Final Remarks

We have defined a principled evolution mechanism that accounts for deployments that correspond to what should be compatible interface evolutions and nevertheless usually force the redeploy (and corresponding downtime) of consumer modules. We address the problem of service interface evolution in a loosely-coupled setting, targeting increased deployment flexibility. We present a language-based lifecycle management model for microservice-based systems that statically checks new service versions, preventing breaking deployments, and certifying the capability of the system to adapt to compatible changes. Also, the system's compatibility algorithm supports several common changes as compatible: adding, reordering and renaming fields. These compatibility scenarios are supported by a runtime adaptation protocol that generates specialised proxies for each producer, ensuring that no communication errors or data loss ever occurs. The flexibility introduced by the referred compatibility relation is framed by the automatic proxy generation. We guaranty that adaptation code is always defined for all pairs of compatible types. The soundness of the type system implies that well defined deployments don't break the system (theorem 1). Our experimental evaluation (section 9) shows significant improvements on the number of independent deployments in real applications that do not require intervention in consumer modules, corresponding to at least 57 % of all deployments with interface changes.

**Acknowledgements** We thank the anonymous reviewers for their comments that helped improving the paper. This work was partially supported by FCT/MCTES grants UID/CEC/04516/2013, PTDC/EEICTP/4293/2014 and PTDC/CCI-INF/32081/2017, and EU H2020 LightKone project (732505).

## About the authors

**João Costa Seco** João Costa Seco is an Assistant Professor at NOVA University of Lisbon. His research interests include language based approaches to security, concurrency, and evolution of software. The main techniques involved are de development of type systems, language semantics, and lightweight software verification tools to be used in real software development environments. Contact him at jrcs@fct.unl.pt

**Paulo Ferreira** Paulo Ferreira is a Lead Software Engineer at OutSystems. He has been a member of the OutSystems R&D team since 2012, where he contributed to the development and research of OutSystems Platform features in the areas of data manipulation, software architecture and API evolution. Contact him at paulo.ferreira@outsystems.com

**Hugo Lourenço** Hugo Lourenço is a Principal Software Engineer at OutSystems. He has been a member of the OutSystems R&D team since 2005, and has contributed with his expertise in many areas of the OutSystems Platform. Lately his focus and responsibilities gravitate around the definition and extensibility of the OutSystems language model. Contact him at hugo.lourenco@outsystems.com

**Carla Ferreira** Carla Ferreira is an Associate Professor at NOVA University of Lisbon. Her research interests are focused on developing formal calculi and tools to express and reason about concurrent and distributed systems, with the overall goal of helping programmers to build trustworthy software systems. Contact her at carla.ferreira@fct.unl.pt

**Lúcio Ferrão** Lúcio Ferrão co-founded OutSystems in 2001 and worked inside OutSystems R&D team until 2018 with multiple technical roles (architecture, compilers, database, scalability, security, language modelling). Since 2018 he works on applied research for OutSystems as an external consultant with a special focus on architecture and compilers. He graduated as a Software Engineer in 1997 at Instituto Superior Técnico from Universidade de Lisboa. Contact him at lucio.ferrao@outsystems.com